\documentclass[12pt]{iopart}

\usepackage{iopams}
\usepackage[left=2.25cm, right=2.25cm, top=2.5cm, bottom=2.5cm]{geometry}
\usepackage{graphicx}								 
\usepackage{amssymb}								 
\usepackage{epstopdf}								 
\usepackage{cite}										 
\usepackage{hyperref}                
\hypersetup{breaklinks=true, colorlinks=true, urlcolor=blue, citecolor=blue} 
\usepackage[all]{hypcap}						 
\usepackage{braket}                  

\newcommand{\td}{\raise.17ex\hbox{$\scriptstyle\sim$}}        
\newcommand{\schrodinger}{Schr\"{o}dinger}										
\newcommand{\lowdin}{L\"{o}wdin}															
\newcommand{\appropto}{\mathrel{\vcenter{
  \offinterlineskip\halign{\hfil$##$\cr
  \propto\cr\noalign{\kern2pt}\sim\cr\noalign{\kern-2pt}}}}}	

\begin{document}

\title{Quantum measurement and thermally assisted proton tunnelling}

\author{A. D. Godbeer, J. S. Al-Khalili and P. D. Stevenson}

\address{University of Surrey, Guildford, Surrey, GU2 7XH, United Kingdom}

\ead{a.godbeer@surrey.ac.uk}

\date{\today \hspace{1 mm}}

\begin{abstract}

Despite compelling evidence to the contrary in recent years, the view still persists that quantum effects cannot survive very long within a warm, noisy and complex environment that washes out quantum effects at timescales far too short for any chemically or biochemically interesting processes. It is also assumed that as the temperature of the surrounding environment increases, so the efficiency of processes such as quantum tunnelling drops. One way of viewing this has been to invoke the quantum Zeno effect: that the watched pot never boils. In this work we show that the opposite is true. For a quite general open quantum system, a proton in an asymmetric double-well potential, the action of the environment is to enhance the tunnelling rate (an anti-Zeno effect). We compare two simple mathematical models to show that, over a specific temperature range, thermally enhanced quantum tunnelling is equivalent to increasing the frequency of a von Neumann-type measurement by the environment on the system.

\end{abstract}


\submitto{\NJP}

\maketitle

\tableofcontents

\section{Introduction}
\markboth{Introduction}{Introduction}
\label{sec:intro}

There are many examples in physics, chemistry and biology of open quantum systems in which the relevant microscopic structures, mechanisms or processes behave quantum mechanically, but which cannot be treated in isolation from their surrounding environment. Typically such quantum systems are embedded within complex molecular structures or are surrounded by water molecules. This external environment is often modelled as a heat bath of harmonic oscillators to describe thermal fluctuations arising from, for example, molecular vibrations. In such open quantum systems, coupling to the environment leads to the loss of quantum coherence at time scales that depend on the temperature of the heat bath and the strength of the coupling. Many different terms are used to describe this process, such as ``relaxation'', ``dissipation'' and ``decoherence''. Open quantum systems are these days a subject of detailed study in situations where the coupling of the quantum system to its environment is a crucial feature of the phenomenon or mechanism of interest, such as in nuclear magnetic resonance. But, by and large, many of the best studied features of the atomic and subatomic world are still dealt with as idealised isolated quantum systems.

In this paper we focus on one mechanism in particular: quantum tunnelling, an example of which is the $\alpha$-decay of an atomic nucleus, where the length scales involved mean the process is treated as a purely quantum mechanical one and the influence of the environment need not be taken into account. There are many cases in chemistry where quantum tunnelling only plays a role when the systems involved are at low temperatures, or when the tunnelling either involves low mass particles, such as electrons, or takes place over short, atomic, distances. However, there exist a number of interesting examples that do not satisfy all these criteria, such as the well-known case of nitrogen inversion in ammonia, whereby a nitrogen atom can tunnel through the plane of the three hydrogen atoms, effectively turning the molecule 'inside out'. It is now also well-established that quantum tunnelling is an important mechanism in electron transfer in proteins whereby single-step tunnelling takes place across distances of up to 10 \AA. 

A well-studied example of proton tunnelling in chemistry has been the double H-bonded benzoic acid dimer \cite{Oppenlander1989, Heuer1991, Meyer_Ernst, Skinner1988, Horsewill1994, Brougham1997, Brougham1999, Horsewill1998, Neumann1998, Neumann1998_2, Horsewill2003}, making this simple chemical system of great use in modelling more complex chemical and biological systems that may involve proton tunnelling.  In biology, one of the first people to speculate on the importance of proton tunnelling was Per-Olov \lowdin~in 1963 \cite{Lowdin1963}, who proposed a model for double-proton tunnelling between H-bonded base pairs in DNA but, despite interest from biochemists over the years \cite{Parker1971, Sokalski1977, Colson1992, Florian1994, Hutter1996, Brameld1997, Bertran1998, Golo2003, Ivanova2011, naturereview}, it has yet to be verified experimentally that proton tunnelling is needed to explain mutations. The first experimental evidence for proton tunnelling in biological systems came in fact from the study of enzyme catalysis in 1989 (for the enzyme alcohol dehydrogenase, which transfers a proton from an alcohol molecule to a molecule of nicotinamide adenine dinucleotide) where the effects of atomic mass on reaction rates through isotopic substitution revealed clear evidence of quantum tunnelling even at relatively high temperatures \cite{Cha1989}. Since then, many other enzymatic reactions have been ascribed to proton tunnelling and it has been established that, at low temperatures, proton tunnelling dominates the proton transfer dynamics\cite{Glickman1994,Masgrau2006,Kohen1999,Scrutton1999}. 

It is now well-established that in a  number of biochemical processes there exists a subtle interplay between quantum coherence and environmental noise such that the action of the latter can assist rather than hinder the former. In this paper, we examine a model of proton tunnelling in a double-well potential under the influence of an external environment. We consider the link between quantum measurement and decoherence using numerical simulations such as has been described by several authors \cite{fearn,Altenmuller1994}, who consider a particle that starts off on one side of a double well potential and investigate the effects of measurement on the time it takes for the particle to tunnel between the wells. However, in those studies conflicting conclusions are reached as to whether continuous measurement slows down the tunnelling process (the quantum Zeno effect) or speeds it up (the anti-Zeno effect) \cite{Kaulakys1997}. For instance, the standard argument is that repeated measurement continually collapses the state of the particle with overwhelming likelihood back to its initial state on one side of the barrier. On the other hand, it is acknowledged that the act of measurement can excite the particle to higher energies and thus enhance the probability of barrier penetration. However, these models are relatively complex and make it difficult to see what is happening in a transparent way. The review by Koshino and Shimizu \cite{Koshino2005} provides a thorough survey of the field but also highlights the complexity of the problem to the extent that the physics can only be appreciated fully by the aficionados.

In particular, a model proposed by Kofman and Kurizki \cite{Kofman2000} is useful for highlighting the advantages of our approach. Their model deals with energy measurements associated with the decay of an unstable state. They derive a universal result they claim shows that the anti-Zeno effect of accelerated decay is much easier to achieve than the Zeno effect itself; the latter being restricted to a limited class of systems due to a competition between the frequency of measurement and the energy spread brought about by this measurement due to the uncertainty principle. Moreover, as the energy uncertainty grows with the frequency of the measurements, the state is able to decay into a larger number of channels, thus accelerating the decay process. Their universal relation involves the convolution of two distributions: the measurement induced level width and spectrum of energies to which the decaying state can couple.

The above model was extended in the work by Ruseckas and Kaulakys \cite{Ruseckas2001,Ruseckas2004}, who take into account both the finite duration and finite accuracy of the measurement. They show that in fact both the QZE and the AZE can be realised depending on the properties of the system and the strength and frequency of the interaction. Just as in Kofman and Kurizki's work, this model also relies on the convolution of the two distributions. When the width of the spectral line (containing the physics of the interaction) is much broader than the width of the reservoir (the range of energy eigenstates available for the decay) the overlap between the two is small and the decay is inhibited (QZE). On the other hand, if the spectral line is narrow then more frequent measurements can broaden it and enhance its overlap with the reservoir spectrum, thus accelerating the decay.

The purpose of this work is to compare the above lines of reasoning with the more traditional approach of dealing with dissipation in an open quantum system using a reduced density matrix model. Here we compare two very different ways of simulating the measurement problem. The first simulates a von Neumann-type process of irreversible ’reduction’ \cite{Neumann1955} via a position (pointer state) measurement that entangles the state of the system with that of the 'measuring device' (in this case the surrounding environment) causing a decay of the off-diagonal elements of the density matrix. The second involves adding a dissipative (Lindblad) term in the master equation for the time-dependant density matrix in order to model the (continuous) coupling of the system to a surrounding heat bath of oscillators. We refer to these two approaches as the pointer and Lindblad methods, respectively. In both approaches, our quantum system consists of a single proton in an asymmetric double well potential whose parameters are chosen to describe a benzoic acid dimer molecule in a crystal field. The two minima along the energy surface for a single benzoic acid dimer are highly symmetric, but introducing a crystal field causes asymmetry, which has previously been determined from temperature-dependent infrared absorptions and NMR data, making one state more energetically favourable than the other \cite{Meyer_Ernst}. Our interest is to ultimately apply these techniques to study proton tunnelling in biological systems, such as in enzyme catalysis and between DNA base pairs, which feature asymmetric well potentials.

\section{Theory}
\markboth{Theory}{Theory}
\label{sec:theory}

\subsection{Pointer measurements}

Many authors have investigated the measurement process in quantum mechanics, which entangles the state of the system with that of the measuring device causing a decay of the off-diagonal elements of the system's density matrix. Wallace \cite{Wallace2001} chose a simple example: the evolution of a one-dimensional wave packet describing the motion of a free particle. He considered the time evolution of the density operator in momentum space then Fourier transformed it back to configuration space before simulating the measurement process by setting all off-block-diagonal elements of the density matrix to zero. He concluded that the effects of repeated measurement can have non-trivial dynamical effects both on the rate of the spreading of the wavepacket and on rate of its centre-of-mass motion, sometimes speeding it up and sometimes slowing it down. But since he only dealt with free particles he could say nothing about the effects of the measurement on quantum tunnelling.

Recently, we extended the idea of Wallace to the more interesting case of the tunnelling of a wavepacket through a square potential barrier \cite{jim_paul} to investigate the more realistic example of particle decay, which is closer in spirit to the models of \cite{misra,fearn,Altenmuller1994}. We describe this approach here and connect it with the more physically realistic reduced density matrix (RDM) approach described in the next section, as well as resolving the dispute as to whether tunnelling is hindered or enhanced through repeated measurement

\begin{figure} [!ht]
    \includegraphics[scale=0.5, clip]{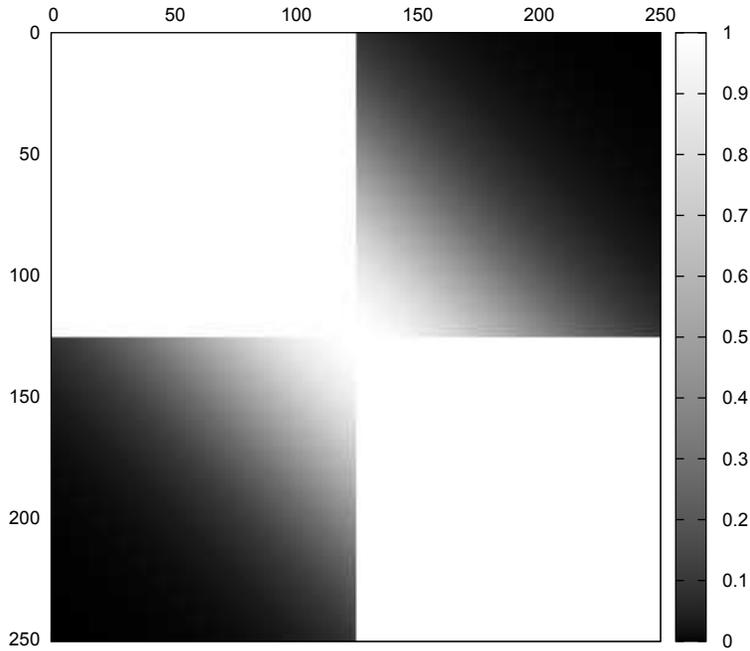}
    \centering
    \caption{The density matrix just after `measurement' showing how the off-diagonal elements in a block of size $N/2$ are removed. White indicates regions of the matrix in which the elements are unchanged and black indicates elements set to zero. Regions shown as grey have been multiplied by a non-zero factor as determined by (\ref{eq:theory_pointer}), with $y = 10^{-4}$.}
    \label{fig:theory_imprecise}
  \end{figure}

We simulate the process of position measurement at any given time for a certain choice of position resolution by setting to zero the off-diagonal terms in the density matrix (expanded in a position space basis). Specifically, we initiate an ``imprecise'' measurement on the system to essentially determine which well the proton is in. As such, only the off-diagonal quarters of the density matrix are set to zero, as shown in Figure~\ref{fig:theory_imprecise}. Unlike other approaches that deal with the coupling of the quantum system to its environment and which have to take into account the number of states available for the decay and the overlap of this (reservoir) spectrum with the measurement induced level width, we do not need to consider the reservoir at all.

We begin by writing the matrix elements of the density operator, $\hat{\rho}(t)= \ket{\psi(t)}\bra{\psi(t)}$
in 1-D coordinate space representation as

  \begin{equation}
    \rho(x,x',t) = \langle x \vert\hat{\rho}(t)\vert x'\rangle =
    \psi(x,t)\psi^\ast (x',t)\ .\label{eq:den2}
  \end{equation}

The time evolution of the density operator of a non-dissipative quantum system is described by the master equation (often also referred to as the quantum Liouville equation)

  \begin{equation}
    \frac{\partial\hat{\rho}}{\partial t} = \frac{1}{\rmi\hbar}[\hat{H},\hat{\rho}]\ ,\label{eq:louiville}
  \end{equation}

where $\hat{H}$ is the Hamiltonian operator for the double well system. We next coarse grain the position into a discrete lattice of position states and expand the system's wave function in a basis of position eigenstates (referred to henceforth as the pointer state basis):  $\ket{\psi(t)} = \sum_{n=1}^{N} C_n (t) \ket{X_n}$ so that all quantities are represented on a grid at uniformly-spaced points $X_n$. Thus, $V_n\equiv V(X_n)$ is the value of the potential at grid point $X_n$. 

Inserting a complete set of states $\sum_k\vert X_k\rangle\langle X_k\vert$ into each term in the commutator in (\ref{eq:louiville}) leads to an equation for the density matrix elements, $\rho_{nm}(t)$ of the form

  \begin{equation}
    \rmi\hbar\frac{\partial\rho_{nm}}{\partial t} = \sum_k(H_{nk}\rho_{km}-\rho_{nk}H_{km}) ,
		\label{eq:eomy}
  \end{equation}

where the Hamiltonian matrix elements are

  \begin{equation}
    \hat{H}_{nm}=\Braket{ n\vert H\vert m} = \left(\frac{-\hbar^2}{2m}
    \frac{\partial\,^2}{\partial X_n^2}+V_n\right)\,\delta_{nm}\ .
    \label{eq:eom2}
  \end{equation}

For ease of computation we then approximate the second derivative in the kinetic energy operator using the three-point formula

  \begin{equation}
    f''(X_n) \approx {\left( f(X_{n-1})-2f(X_n)+f(X_{n+1}) \right) / \Delta X^2}\ ,
		\label{eq:3pt}
  \end{equation}

where $\Delta X$ is the grid spacing. (\ref{eq:eom2}) then simplifies through the use of (\ref{eq:3pt}) and the fact that the potential is diagonal, to

  \begin{equation}
    \dot{\rho}_{nm} = \frac{1}{\rmi \hbar} \left[ - \frac{\hbar^2}{2m \Delta X^2} \left( \rho_{n-1,m} + \rho_{n+1,m} - \rho_{n,m-1}-\rho_{n,m+1}\right) + (V_n-V_m)\rho_{nm} \right],
		\label{eq:eom3}
  \end{equation}

The above first order differential equation in time is solved using Runge-Kutta algorithm for the full $N\times N$ coupled equations (one for each element in the density matrix).

Simulating a pointer measurement is relatively simple. First, a \textit{block size} (from anywhere between 1 and $N/2$, where $N$ is the number of pointer states) is chosen to determine the ``precision'' of the measurement. In this study, we choose a block size of $N/2$, which effectively allows us to know, upon measurement, no more than which side of the barrier the proton is on (an imprecise measurement). At certain chosen intervals, greater than the step size in time in the numerical integration of (\ref{eq:eom3}), we carry out our 'measurement' by multiplying all elements in the off-diagonal blocks of the density matrix by a decay factor

  \begin{equation}
    {\rho}_{nm} = {\rho}_{nm} \rme^{-y {\left( n - m \right)}^2}\ ,
		\label{eq:theory_pointer}
  \end{equation}
where $y \geq 0$ is a `harshness' parameter.

It is clear from (\ref{eq:theory_pointer}) that the closer to the diagonal the element is ($n-m$ is small) the less affected it is, but for those further away ($n-m$ is large) the closer to zero the decay factor is, effectively `killing off' the matrix element, as shown in Figure~\ref{fig:theory_imprecise}.
  
\subsection{The Lindblad method}

The density matrix can also be expanded in the basis of energy eigenstate of the double well, where a smaller matrix is required (since the number of eigenstates needed to describe the wave function is far fewer than the number of grid points considered in the pointer basis). In this basis the density matrix is
  
  \begin{equation}
    \label{eq:theory_master_part12}
      \hat{\rho} = \ket{\psi} \bra{\psi} = \sum\limits_{ij}^{} \alpha_i \alpha_j^* \ket{\phi_i} \bra{\phi_j} , \qquad {\rm where,}\ \ \  \hat{H} \ket{\phi_i} = E_i \ket{\phi_i} ,
  \end{equation}

which gives a particularly simple form for the time evolution of the density matrix elements
  
  \begin{equation}
      \dot{\rho}_{ij} = \frac{1}{\rmi \hbar}\left( E_i - E_j \right) \rho_{ij} .
  \end{equation}

We will discuss later on the number of eigenstates necessary in order to achieve good enough accuracy.

In order to model an open (dissipative) system, coupling to the environment (in the limit of weak coupling to a Markovian bath) can be included in the Liouville Equation~(\ref{eq:louiville}), which is generalised to include a dissipative (Lindblad) term on the right hand side \cite{Gorini1976,Lindblad1976}

  \begin{equation}
    \label{eq:theory_master_repeat2}
      \frac{\partial \hat{\rho}}{\partial t} = \frac{1}{\rmi \hbar} [\hat{H}, \hat{\rho}] + \hat{L} \hat{\rho} 
  \end{equation}

where this extra term generally takes the form
  
  \begin{equation}
    \label{eq:theory_master_part20}
      \hat{L} \hat{\rho}  = \sum\limits_{\beta}^{} \left( \hat{A}_{\beta} \hat{\rho} \hat{A}_{\beta}^{\dagger} - \frac{1}{2} \left[ \hat{A}_{\beta}^{\dagger} \hat{A}_{\beta}, \hat{\rho} \right]_+ \right) 
  \end{equation}

and the operators $\hat{A}_\beta$ are defined as \cite{scheurer_saalfrank} 

  \begin{equation}
    \label{eq:theory_master_part21}
      \hat{A}_{\beta} = \sqrt{W_{ij}} \ket{i} \bra{j}\ ,
  \end{equation}
  
The index $\beta$ labels ordered pairs $(i,j)$ of energy eigenstates and $W_{ij}$ are environment induced transition rates between well states $\ket{i} and \ket{j}$.  This leads to
  
  \begin{equation}
    \label{eq:theory_master_part29}
      \hat{L} \hat{\rho} = \sum\limits_{ij}^{} W_{ij} \left( \ket{i} \bra{j} \hat{\rho} \ket{j} \bra{i} - \frac{1}{2} \left[ \ket{j}\bra{j} , \hat{\rho} \right]_+ \right).
  \end{equation}

Substituting the above back into (\ref{eq:theory_master_repeat2}) leads to diagonal and off-diagonal density matrix elements in the eigenstate basis 
 
  \begin{eqnarray}
    \label{eq:drho_dt_W_jk}
      \dot{\rho}_{ij} &=& \frac{1}{\rmi \hbar} \left( E_i - E_j \right) \rho_{ij} - \frac{1}{2}\ \rho_{ij} \sum\limits_{k}^{} \left( W_{ki} + W_{kj} \right) , \quad \quad \quad i \neq j \nonumber \\ \nonumber \\
      \dot{\rho}_{ii} &=& \sum\limits_{k}^{} \left( W_{ik} \rho_{kk} \right) - \rho_{ii} \sum\limits_{k}^{} \left( W_{ki} \right) .
  \end{eqnarray}

There are a number of ways of calculating the transition (relaxation) matrix elements, $W_{ij}$. They are derived here using the microscopic theory of Meyer and Ernst \cite{Meyer_Ernst}. We will not repeat the details of the derivation, but will summarise the assumptions and inputs to the model briefly. The full system+bath Hamiltonian is written as

  \begin{equation}
    \label{eq:results_H_parts}
      \hat{H} = \hat{H}_s + \hat{H}_b + \Delta{\hat{H}}\ ,
  \end{equation}
  
where $\hat{H}_s$ is the system Hamiltonian involving a kinetic energy operator for the tunnelling particle along with the double well potential and $\hat{H}_b$ is the bath Hamiltonian defined as a sum of harmonic oscillators 

    \begin{equation}
    \label{eq:results_H_environment}
      \hat{H_b} = \frac{1}{2}\sum\limits_{m}^{} \left( p^2_m + \omega^2_m q^2_m \right),
  \end{equation}

where $m$ is the set of bath oscillators, $p_m$ are their momenta, $q_m$ are their spatial positions and $\omega_m$ their frequency. Finally, $\Delta \hat{H} = \sum\limits_{m}^{} f_m q_m$ is the interaction between the system and bath, with coupling constant, $f_m$.

The transition probability $W_{jk}$ between states $j$ and $k$ is defined as

  \begin{equation}
    \label{eq:results_W_jk}
      W_{jk} = \frac{1}{\hbar^2} \int\limits_{- \infty}^{\infty} \rmd \tau \rme^{-i \omega_{jk} \tau} C_{jk}(\tau)\ ,
  \end{equation}
  \begin{equation}
    \label{eq:results_W_kk}
      W_{kk} = - \sum\limits_{j \neq k}^{} W_{jk},
  \end{equation}

where $\omega_{jk}$ is a transition frequency depending on the energy of the eigenstates $j$ and $k$:

  \begin{equation}
    \label{eq:results_w_jk}
      \omega_{jk} = \frac{\left(E_j - E_k \right)}{\hbar}
  \end{equation}

and the transition probabilities, $W_{jk}$, according to this definition will automatically fulfil the principle of detailed balance \cite{Meyer_Ernst}.

The correlation functions, $C_{jk}$, required in (\ref{eq:results_W_jk}) are calculated from an appropriately chosen power spectral density function of the active bath displacement:

  \begin{equation}
    \label{eq:results_J_rr}
      J_{rr}(\omega) = \frac{4 \sqrt{2} \Delta V_R \hbar \omega_p \omega^3}{\left(\omega_p^4 + \omega^4 \right) \left(\rme^{\frac{\hbar \omega}{k_B T}} -1 \right)},
  \end{equation}

where $T$ is temperature, $\omega_p$ is a characteristic phonon frequency of the heat bath and $\Delta V_R$ is the rearrangement energy gained by the bath oscillators upon displacement from $q_m$ = 0 to their optimal values at the potential minima \cite{Meyer_Ernst}. This definition of the power spectral density function is related to the chosen model for the bath oscillators, using Debye theory, whereby the product of the square of the coupling constants and the density of modes increases with $\omega^4$ at low frequencies and becomes constant at $\omega_p$.

\section{Numerical Results}
\markboth{Numerical Results}{Numerical Results}
\label{sec:results}

  \begin{figure}
    \hbox{\hspace{16ex}\includegraphics[scale=0.48, clip]{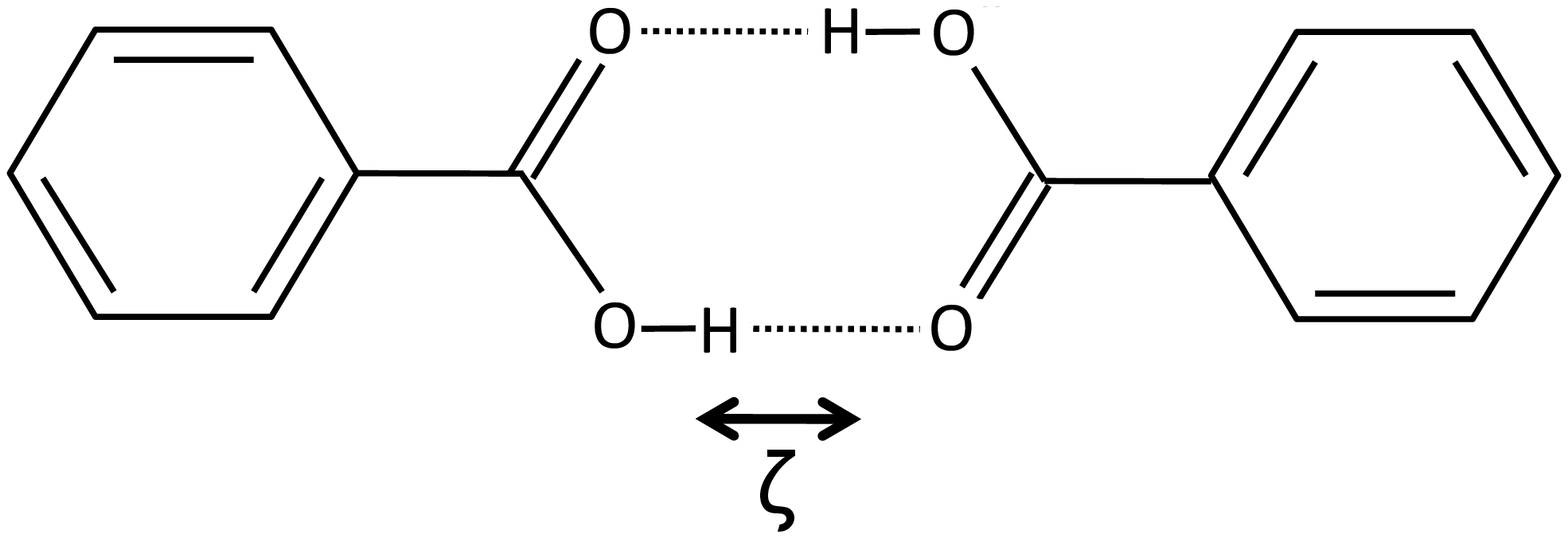}}
		\includegraphics[scale=0.5, clip]{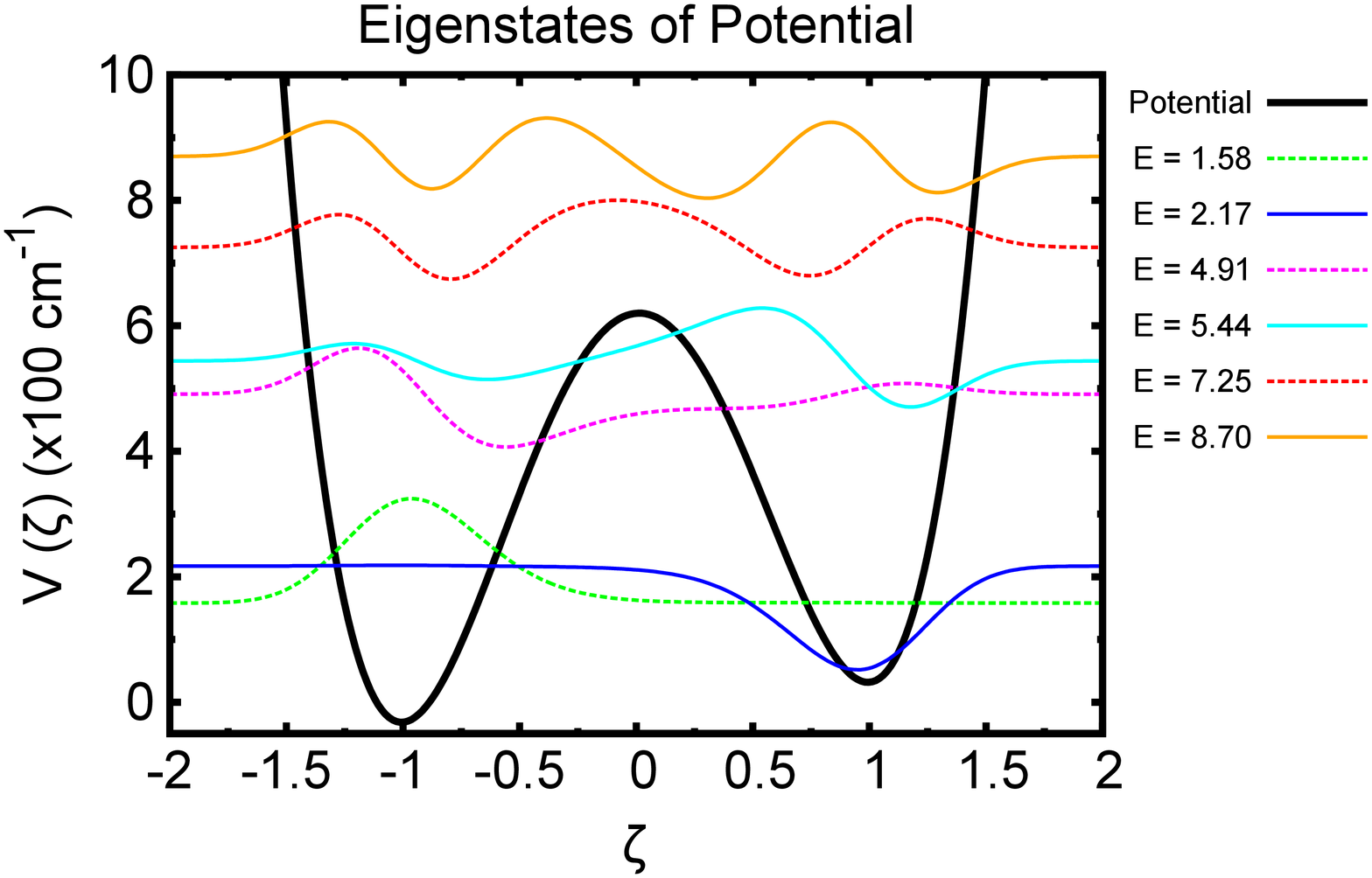}
    \centering
    \caption{Double well potential for hydrogen transfer in benzoic acid dimer, showing the first 8 energy eigenstates. States 1 \& 2 are wholly localised in their respective wells, and states 3 \& 4 are mostly localised in their respective wells.}
    \label{fig:result_well_eigenstates}
  \end{figure}
	
We use an analytic one-dimensional asymmetric double well potential with a quartic dependence on position \cite{Meyer_Ernst}:
  
  \begin{equation}
    \label{eq:potential}
      V(\zeta) = \left(B \left(1 - \zeta^2 \right) \right)^2 + \frac{\Delta V }{2} \zeta,
  \end{equation}

where $B$ is the barrier height, $\Delta V$ is the asymmetry parameter and the dimensionless variable $\zeta$ is a reduced proton position transfer co-ordinate. The parameters were chosen, as in \cite{Meyer_Ernst}, to describe the benzoic acid dimer: $B$ = 620 cm$^{-1}$,  $\Delta V$ = 63.6 cm$^{-1}$. Figure~(\ref{fig:result_well_eigenstates}) shows the chemical structure of the benzoic acid dimer, with its double H-bond, and the shape of the resulting double well potential. Also plotted are the first six eigenfunctions and their corresponding energies, calculated from a numerical solution of the \schrodinger~equation using the a Runge-Kutta routine.

As a test of the two approaches, we first carry out calculations for the isolated system. That is, without setting the off-diagonal elements of the density matrix to zero in the pointer method and without the Lindblad term in the eigenstate basis (achieved by setting the temperature $T=0$ in the spectral function, $J$). An initial Gaussian wave function, centred at $\zeta = -1$, was chosen to represent a proton that starts off at $t=0$ in the deeper well. The density matrix was then evolved in time in each case and the probability of finding the proton in the shallow well calculated at appropriate time steps. Due to the asymmetry of the double well, the ground state eigenfunction exists mainly in the deeper well and the initial Gaussian wave function consists almost entirely of this lowest eigenfunction. Crucially of course, it will contain small components of higher eigenstates and is thus not a stationary state. A small component of the wave function will therefore tunnel through to the shallow well.
  
\begin{figure} [tb]
    \includegraphics[scale=0.4, clip]{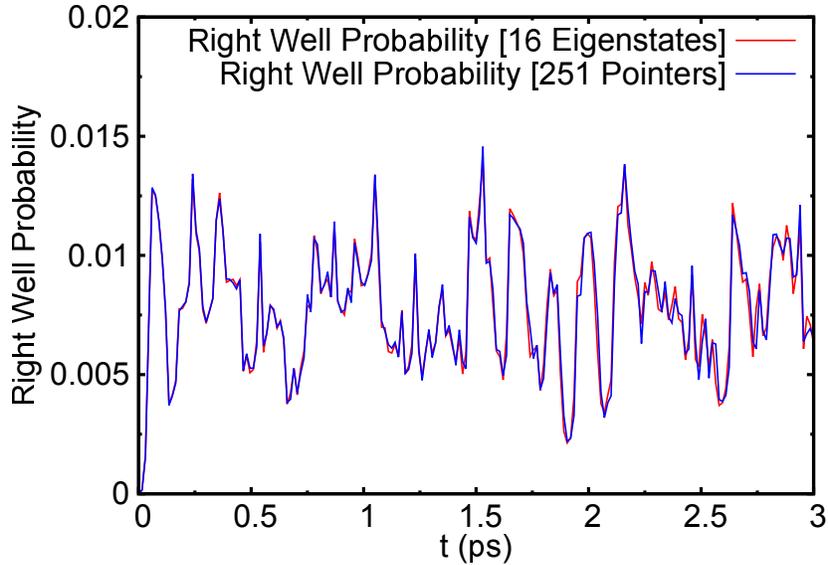}
    \centering
    \caption{Probability of proton being in shallow well over a period of 3 ps using optimum basis sizes with no measurement/environment coupling . The results match extremely well, demonstrating the accuracy of the two quite different approaches.}
    \label{fig:result_basisset_comparison}
  \end{figure}

In the absence of any interaction with the environment, the numerical accuracy of the two methods can be tested against each other to determine the sizes of the basis sets required for sufficient accuracy. It was found that the eigenstate basis required 16 terms, which produces results matching those obtained in the pointer state basis to within an accuracy of one part in $10^5$ over the whole time period of interest. Although rather difficult to see, there are in fact two curves in Figure~(\ref{fig:result_basisset_comparison}). The low probability of tunnelling reflects the choice of initial Gaussian wave function and we stress the agreement between the two approaches is no more than a test of the numerical calculations.

We next include the coupling to the environment by a) adding the dissipative \textit{Lindblad term} in the eigenstate basis and b) carrying out pointer measurements at regular intervals. Instead of taking a Gaussian centred in the deeper well for the initial wave function (depicting a proton that is definitely in the deeper well to begin with), we now choose the more realistic case of the ground state eigenfunction, $\ket{\psi(t=0)}=\ket{\phi_0}$. Due to the asymmetry of the well, this eigenfunction is almost entirely in the deep well anyway and closely resembles the Gaussian shape. However, this is a stationary state and any tunnelling will now be due entirely to the coupling of the system to the environment. In the Lindblad method, this is due to the external heat bath inducing transitions between the ground state and higher energy eigenstates that have larger components in the shallow side of the well (which is just another way of saying it enhances the tunnelling probability). Since this coupling depends on the temperature of the bath we should expect an enhanced tunnelling probability with higher temperature (\textit{thermally assisted tunnelling}). In contrast, the interpretation in the pointer method is that repeated measurement (setting the off-diagonals of the density matrix to zero) provides an unavoidable 'kick' to the proton pushing it to higher energy states.

Clearly, if the coupling is strong enough in both pictures the proton could be induced into hopping over the barrier classically rather than tunnelling through it. To test this, we compared two eigenstate basis calculations with basis sizes of 4 and 16 eigenstates, respectively. The first four energy eigenvalues lie below the top of the barrier and so, with a basis that only includes coupling between these states, quantum tunnelling would be the only way for the proton to find itself on the other side of the barrier. On the other hand, including a further 12 states would allow classical 'over the barrier' hopping. Hardly any difference at all was found between the two cases, implying that tunnelling is the dominant mechanism in this case.

In addition, the occupation probability of each eigenstate (obtained by finding the overlap between the wavefunction and each eigenstate) over time was checked for the two cases of basis size.

  \begin{figure}
    \includegraphics[scale=0.4, clip]{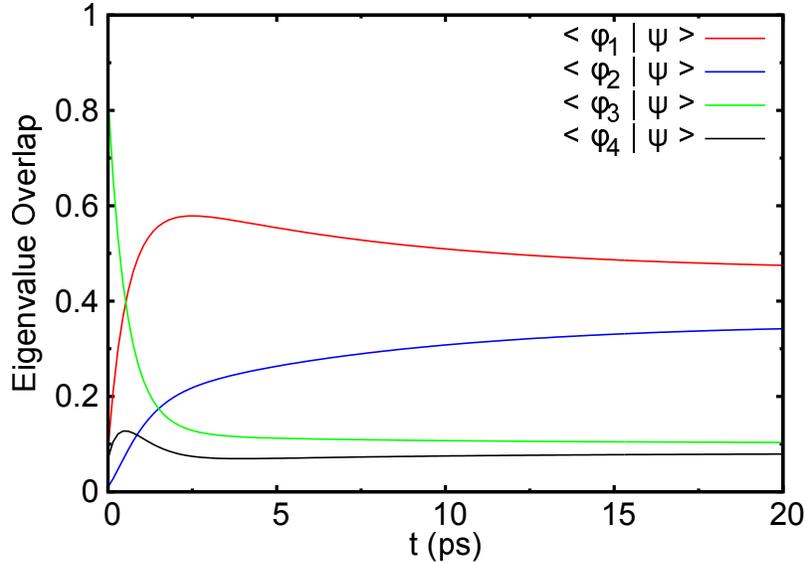}
    \centering
    \caption{Overlap of the wave function with each of the first four energy eigenstates over a period of 20 ps.}
    \label{fig:result_4_states_overlap}
  \end{figure}

  \begin{figure}
    \includegraphics[scale=0.4, clip]{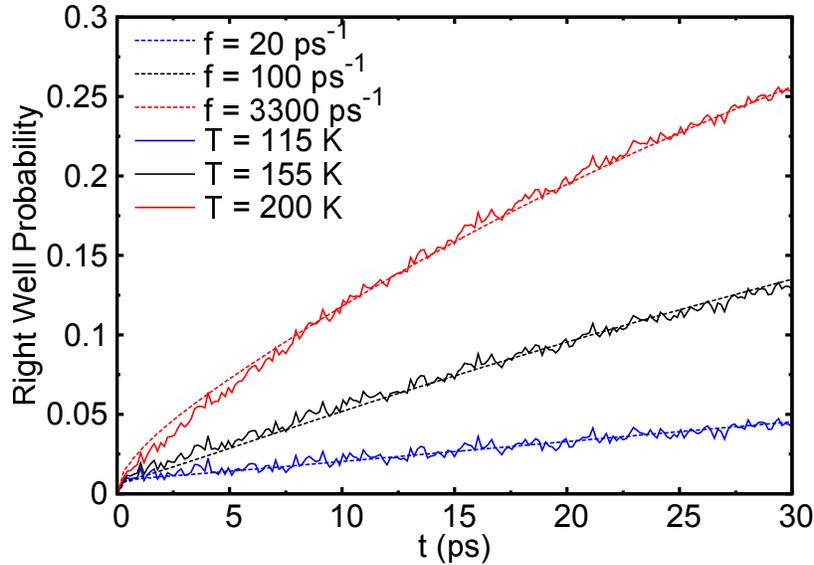}
    \centering
    \caption{Probability of proton being in shallow well during a period of 30 ps using pointer state measurements with varying observation frequencies (f) and the Lindblad term with varying temperatures (T). Observation size is fixed at 0.5N and harshness is set at $10^{-4}$.}
    \label{fig:result_well_probabilities_comparison}
  \end{figure}
	
Figure~(\ref{fig:result_4_states_overlap}) shows how the occupation probability for first four eigenstates changes over time. When using the larger basis, the higher 12 energy eigenstates have a cumulative occupation probability of 6\% at 20 ps. What is clear is that adding the dissipative Lindblad term allows primarily for strong transitions between the ground state and the first excited state (which is predominantly in the shallow well). Thus, the thermally assisted tunnelling is almost entirely due to populating this state.

Figure~(\ref{fig:result_well_probabilities_comparison}) shows a comparison between two sets of curves. The solid lines are from the eigenstate basis calculation at three different bath temperatures (115 K, 155 K, and 200 K). Overlaying these are the results (dashed lines) from the pointer calculation with measurements being made at three different frequencies of 20 ps$^{-1}$, 100 ps$^{-1}$, and 3300 ps$^{-1}$. (Note the numerical step size in $t$ is 0.05 fs.) It is clear from this graph that, on the one hand, increasing the temperature of the bath leads to stronger coupling to the environment and hence enhanced rate of thermally assisted tunnelling. This is very strongly and neatly correlated with a similar enhancement in tunnelling probability with increased frequency of observation/measurement - what might be referred to as an 'anti-Zeno effect'. Making the link in this way with dissipative Lindblad approach clarifies why this is so.

  \begin{figure} [!ht]
    \includegraphics[scale=0.46, clip]{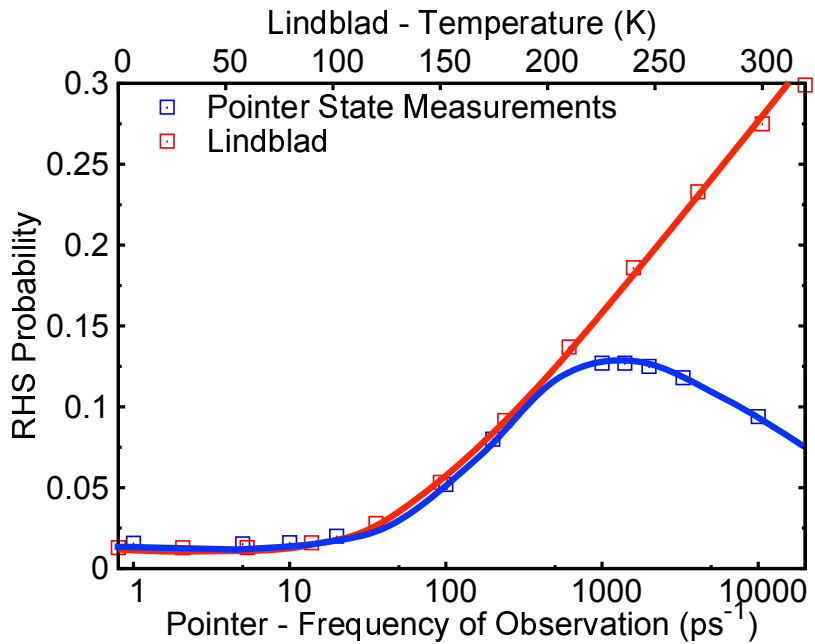}
    \centering
    \caption{Probability of the proton being in the shallow well after 10 ps using pointer
state measurements with varying observation frequencies and the Lindblad approach with varying temperatures. The sold curves have been drawn through the discrete calculated points to guide the eye.}
    \label{fig:result_well_probabilities_10ps}
  \end{figure}

  \begin{figure} [!ht]
    \includegraphics[scale=0.46, clip]{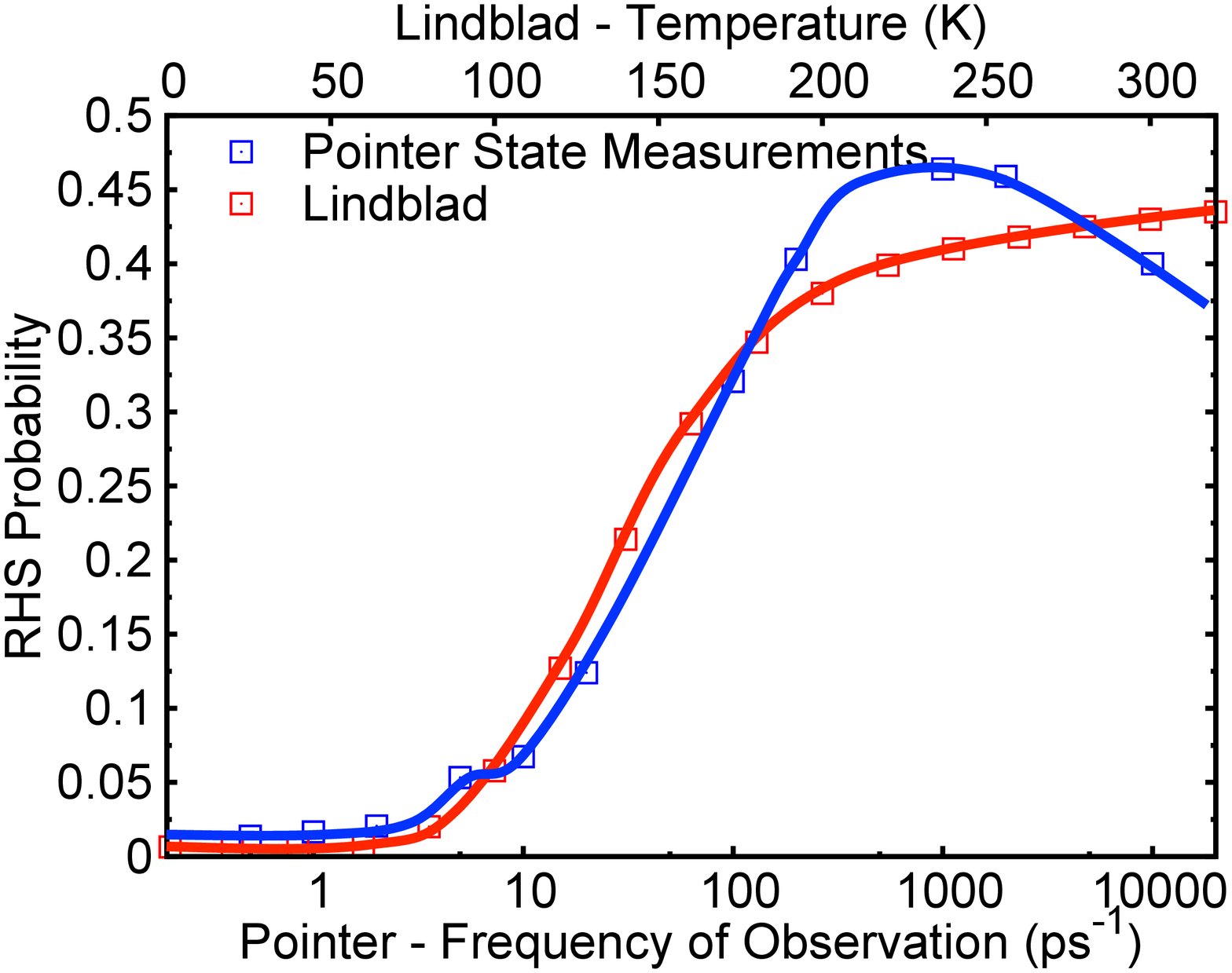}
    \centering
    \caption{Probability of the proton being in the shallow well after 100 ps using pointer measurements with varying observation frequencies and the Lindblad approach with varying temperatures. The sold curves have been drawn through the discrete calculated points to guide the eye.}
    \label{fig:result_well_probabilities_100ps}
  \end{figure}

Figures~(\ref{fig:result_well_probabilities_10ps}) and (\ref{fig:result_well_probabilities_100ps}) consider two different snapshots in time: at $t=10$ ps and $t=100$ ps, and compares the probability of the proton having tunnelled as a function of temperature in the Lindblad approach with its probability of having tunnelled as a function of observation frequency in the pointer approach. A similar picture is seen at both times: increasing the frequency of measurement by the environment is equivalent to raising its temperature -- both lead to enhanced tunnelling probability, or anti-Zeno effect.

Beyond 200 K in the Lindblad approach, the tunnelling probability continues to rise until it reaches a maximum of $P\leq 0.5$, depending on the asymmetry in the double well. Figure~(\ref{fig:result_4_states_overlap}) explains the reason for this levelling off since, after about 20 ps, the occupation probabilities of each energy eigenstate remain constant. On the other hand, increasing the measurement frequency in the pointer model is equivalent to imparting a 'kick' to the proton, exciting to higher energy states where it can tunnel more easily. But when the measurements are made too frequently (beyond 1000 ps$^{-1}$) we see a drop in tunnelling probability and a clear change from anti-Zeno to Zeno effect. This can be interpreted as the wave function being collapsed back to its initial state (by setting the off-diagonals of the density matrix to zero) so often that it does not have as much chance to evolve and the proton is less likely to tunnel. 
  
It is clear that the pointer and Lindblad methods agree in their simulation of an environment acting upon the system for temperatures up to around 200 K. Beyond this, the methods' predictions drift apart.  This correspondence suggests that quantum measurements may effectively drive a system to a given virtual temperature, and may prove useful in explaining how, if it is indeed the case, quantum effects can persist in biological systems in an environment when one might rather expect them to be washed out.

\section{Conclusions}
\markboth{Conclusions}{Conclusions}
\label{sec:conclusions}

With an asymmetrical double well potential, such as can be found in many chemical and biological systems, including the benzoic acid dimer modelled here, we have shown that two very different pictures of environment induced decoherence can be compared with each other, and a clear and simple link made between the frequency of von Neumann type measurement and the temperature of a dissipative environment (heat bath). We conclude that, for the simple model described here, increasing the strength of coupling to the environment (achieved by raising the temperature of the heat bath) leads to a clear anti-Zeno effect of enhancing the tunnelling rate. This rate continues to rise over time, reaching a maximum (that depends on the asymmetry of the well) by around 100 ps. Up to a temperature of 200 K this enhanced tunnelling can be mimicked very well by increasing the frequency of a von Neumann type measurement. However, increasing the frequency of measurement further, corresponding to a bath temperature of over 200 K, leads to a changeover from anti-Zeno to Zeno effect whereby the tunnelling rate starts to decrease again -- in contradiction to the Lindblad approach where increasing the temperature further (from 200 K to room temperature) continues to enhance the tunnelling rate until it reaches a stable plateau. We have also shown that this thermally assisted tunnelling is a general feature of such a double-well system as it couples lower energy states to higher ones closer to the top of the barrier. A similar picture can be invoked in the pointer approach whereby frequent measurement will disturb the system and excite the proton. And while very frequent measurements do show a Zeno effect (of collapse of the quantum state to its initial, untunnelled, one), this is not borne out in the more realistic Lindblad picture.

\section{Acknowledgements}

The authors would like to thank the Engineering and Physical Sciences Research Council (EPSRC) for funding the Doctoral Training Centre and this work.

\pagebreak
\markboth{Bibliography}{Bibliography}
\bibliographystyle{iopart-num}
\bibliography{ADG_JAK_PDS}

\end{document}